# Galaxy Zoo: Motivations of Citizen Scientists

M. Jordan Raddick
*Johns Hopkins University, Baltimore, Maryland, 21210*

Georgia Bracey
*Southern Illinois University Edwardsville, Edwardsville, Illinois, 62026*

Pamela L. Gay
*Southern Illinois University Edwardsville, Edwardsville, Illinois, 62026*

Chris J. Lintott
*Oxford University, Oxford, OX1 3RH, United Kingdom*

Carie Cardamone
*Brown University, Providence, RI, 02906*

Phil Murray
*Fingerprint Digital Media, Newtownards, Northern Ireland, BT23 7GY, United Kingdom*

Kevin Schawinski
*Eidgenössische Technische Hochschule (ETH), 8093 Zürich, Switzerland*

Alexander S. Szalay
*Johns Hopkins University, Baltimore, Maryland, 21210*

Jan Vandenberg
*Johns Hopkins University, Baltimore, Maryland, 21210*

ABSTRACT

Citizen science, in which volunteers work with professional scientists to conduct research, is expanding due to large online datasets. To plan projects, it is important to understand volunteers' motivations for participating. This paper analyzes results from an online survey of nearly 11,000 volunteers in Galaxy Zoo, an astronomy citizen science project. Results show that volunteers' primary motivation is a desire to contribute to scientific research. We encourage other citizen science projects to study the motivations of their volunteers, to see whether and how these results may be generalized to inform the field of citizen science.

Interdisciplinary
Research into teaching/learning
Web-based learning
Amateur astronomers and education
OTHER: Citizen science
OTHER: Volunteers
OTHER: Motivation

## 1. INTRODUCTION

"Citizen science" is a scientific practice in which volunteers from the general public assist scientists in conducting research. Citizen scientists have worked with professional scientists to collect data for centuries (Silvertown 2009, Bracey 2010). Today, the presence of massive online scientific datasets and the easy availability of high-speed Internet access provide new opportunities for citizen scientists to work on projects analyzing and interpreting existing datasets. Examples of this new form of citizen science in astronomy are detailed in Mendez (2008).

This new method of citizen science has shown great promise in advancing both scientific research and public understanding of science – but developing successful citizen science projects requires a clear understanding of citizen science from the volunteers' perspective. Why do volunteers choose to contribute their time and energy to citizen science? What do they get out of it?

Although there is a wide scholarly literature on motivations for volunteering, with a variety of perspectives and methods (e.g. Clary & Snyder 1999, Shye 2010), there has been little empirical research exploring motivations for volunteering for citizen science activities specifically (Nov, Arazy, & Anderson 2011). Because citizen scientists may have different motivations for volunteering than people who participate in other volunteer activities, it is important to study motivation to volunteer specifically in the context of citizen science. This paper is the second in a series that attempts to explore these important questions.

In this paper, we report results of a survey to measure the motivations of volunteers participating in a successful online data analysis citizen science project in astronomy. "Galaxy Zoo" invites volunteers to classify the shapes of galaxies seen in images of the Sloan Digital Sky Survey. For more information about Galaxy Zoo, including a discussion of how volunteers' galaxy classifications compare to professionals', see Lintott, et al. (2008).

Motivation has been defined differently in a variety of contexts and studies; for our current purpose, we consider motivation to be a mental construct that a volunteer uses, consciously or unconsciously, to explain their behavior, arising out of a combination of the person's mental state and properties of the situation they are in (Humphreys & Revelle 1984). A number of different theoretical frameworks can be used to analyze motivation; we consider some of these frameworks in light of our results in our discussion in section 6.2.

In our previous research (Raddick et al. 2010), we conducted open-ended interviews with Galaxy Zoo volunteers, which analyzed using methods from grounded theory (Strauss & Corbin 1990), to identify twelve motivations for participation in Galaxy Zoo. Table 1 shows the motivations identified in that research, along with the stem used to describe the motivation to volunteers in the survey instrument whose results are reported here.

| **Motivation** | **Description (used in survey instrument)** |
|---|---|
| Contribute | I am excited to contribute to original scientific research. |
| Learning | I find the site and forums helpful in learning about astronomy. |
| Discovery | I can look at galaxies that few people have seen before. |
| Community | I can meet other people with similar interests. |
| Teaching | I find Galaxy Zoo to be a useful resource for teaching other people. |
| Beauty | I enjoy looking at the beautiful galaxy images. |
| Fun | I had a lot of fun categorizing the galaxies. |
| Vastness | I am amazed by the vast scale of the universe. |
| Helping | I am happy to help. |
| Zoo | I am interested in the Galaxy Zoo project. |
| Astronomy | I am interested in astronomy. |
| Science | I am interested in science. |

Table 1. Categories of motivation used in this research. The table lists the name used by the research team to describe the category, and the description of the category that was used as a stem in the survey instrument.

In this paper, we report results of an online follow-up survey that asked Galaxy Zoo volunteers to self-report their motivations using the same twelve categories, with an option to enter other motivations as free responses. We describe the survey instrument and its implementation in section 2, and the sample of responses in section 3. Section 4 reports results of demographic questions on the instrument, while section 5 reports results of motivation questions. Section 6 discusses and interprets these results. Section 7 outlines potential future work that could uncover other aspects of motivation in Galaxy Zoo and other citizen science projects; section 8 considers potential larger implications if these results can be generalized to the field of citizen science.

## 2. THE GALAXY ZOO MOTIVATIONS SURVEY

Our prior research resulted in a list of twelve possible motivations of Galaxy Zoo volunteers (Raddick et al. 2010); these motivations are shown in Table 1. The methods we used in this prior work allowed us to conclude that these motivations were present to some extent in our population, but did not allow for an estimate of the prevalence of any of them – such an estimate would require asking motivation questions of a much larger sample of volunteers using a survey.

To estimate the prevalence of our twelve possible motivations, as well as how those motivations may vary with demographic characteristics, we conducted an online survey. Our survey instrument is described in section 2.1, and the implementation of the survey is described in section 2.2.

### 2.1. Survey Instrument

The complete list of the survey items we used is given in Appendix A. To collect demographic data, we asked participants to report their age, gender, country of residence, and level of education. Because Galaxy Zoo is an international project, we gave volunteers the choice to report their education levels using either U.S.-style or U.K.-style education systems.

Our survey instrument asked about volunteers' motivations in three ways. First, because the interviews we conducted previously suggested that most volunteers reported multiple motivations for participating, we gave respondents the opportunity to self-report their perceptions of each of the twelve motivations independently. For each, we asked, "How motivating are these reasons to you?" Responses were on a seven-point Likert scale (Likert 1932), with endpoints "not motivating to me" and "very motivating to me."

Second, immediately after, we asked, "Can you think of any other reasons that someone might be interested in Galaxy Zoo? If so, list them here." This question gave respondents the ability to report motivations other than those in our categorization scheme. We hypothesized that that phrasing the question to elicit perceptions of others' possible motivations might encourage respondents to report a wider range of possible motivations than they might have first considered.

Third, to identify volunteers' primary motivation for participating, we asked: "Which of the reasons above is MOST important to you for participating in Galaxy Zoo? Select the radio button next to your most important reason." The survey instrument listed each of the twelve motivations, and survey-takers clicked a radio button to indicate their preference. An "other" option was provided. This question allowed us to place all volunteers into groups based on motivation, which could then be analyzed separately and compared.

### 2.2. Survey Implementation

We chose to implement the survey online; since Galaxy Zoo is an entirely online project, we reasoned that offering the survey online would not lead to any response bias, while the ease of automatically recording responses in a database would be a large advantage. We implemented the survey as a web form with a back-end MySQL database, hosted on the Astrosphere New Media Association website (astrosphere.org); we included a link to the survey in our messages to potential survey-takers. A screenshot of the online survey implementation is shown in Figure 1.

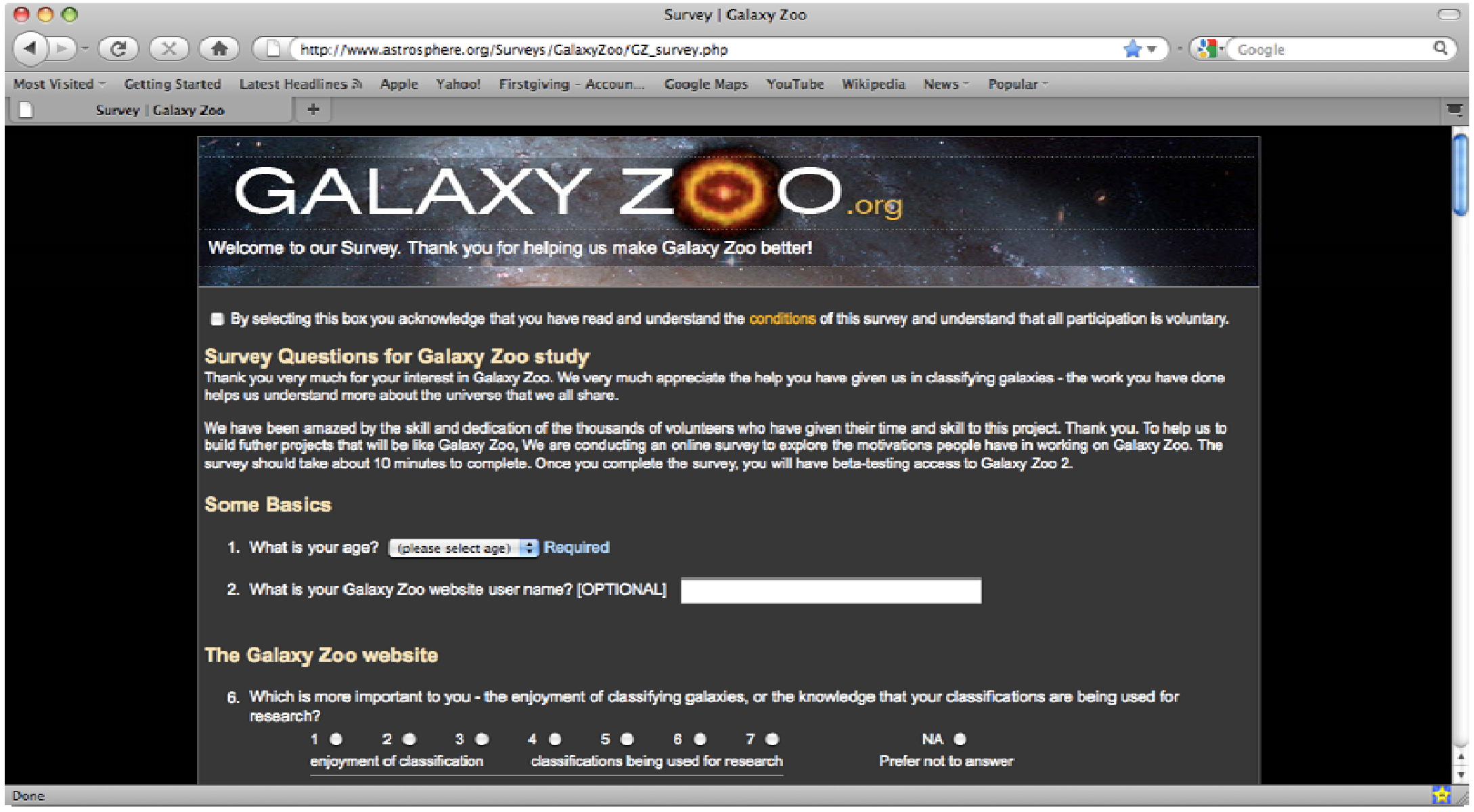

Figure 1. A screenshot of the online survey form.

When a participant arrived at the web form, they were randomly assigned one of two versions of the survey instrument. One version had demographic questions at the beginning of the instrument, the other at the end. We did this to investigate the possibility of stereotype threat effects (Spencer, Steele, & Quinn 1999) – the chance that some volunteers could be intimidated by being asked up front for their gender or education level, and might therefore self-report different motivations.

We implemented the survey in two phases. The first phase featured an incentive for completing the survey: early beta access to the Galaxy Zoo 2 site (a link to the beta site was provided at the end of the survey). Although there is some risk of selection bias due to the fact that the incentive is closely tied to motivation, previous experience has shown that for online astronomy projects, a content-based incentive maximizes participation (Gay et al. 2007).

The first phase began on November 20th, 2008. On that date, we sent out a survey solicitation as part of the regular Galaxy Zoo e-newsletter. Although we do not have a record of how many people received this email, the total number of Galaxy Zoo users on that date can be used as a proxy for this number, for use in estimating the survey response rate. At the time, there were 174,764 volunteers who had registered with Galaxy Zoo; nearly all received the newsletter. The first phase of the survey implementation had received 10,991 responses – giving an estimated response rate of 6.3%. Because not every Galaxy Zoo volunteer signed up for the newsletter using a valid email address, the actual number of people who received the solicitation must have been lower – so the 6.3% response rate should be considered a lower limit only.

Later, we advertised the survey again, through the Galaxy Zoo forum, and collected responses from this second survey implementation in a separate database table. In this survey implementation, no incentive was offered. This second survey implementation received 871 responses. Because the second survey implementation was advertised primarily through the Galaxy Zoo forum, responses come from a sample of the population of users on the Galaxy Zoo forum. The population of forum users is much smaller than the total population of Galaxy Zoo volunteers, and it is likely that motivations present in this population differ from those of the population of all Galaxy Zoo volunteers. That difference, the absence of an incentive for participation, and the very different sample sizes preclude straightforward comparison of motivations between the two samples and incentive cases.

We did not have an official cutoff date for survey responses, but we downloaded our final response dataset on March 24th, 2009. No responses submitted after that date are included in this paper. Our raw dataset contained 11,862 total survey responses. We describe the process of cleaning the raw dataset in section 3.1.

## 3. SAMPLE

### 3.1. Data Cleaning

Before analyzing our data, we searched for and removed potentially problematic responses in three ways.

First, in keeping with our study's IRB requirements, we removed responses from participants who self-reported their ages as less than 18.

Next, we removed potentially duplicate responses; we defined duplicates as two or more responses from the same IP address with identical responses to demographic questions. Within each pair of duplicates, the record with most data was retained; if the two had the same amount of data, we kept the first (earliest) response.

To account for the possibility that some responses came from participants rushing through the survey, we checked whether responses to motivation questions using the Likert scale had all the same value (e.g. all 1s or 4s). Some such responses existed, but visual inspection of those records suggested that most represented genuine attempts to answer accurately. This suggested that, rather than try to judge which specific responses were or were not valid, we should keep them all.

After removing responses with ages less than 18 and likely duplicate responses as described above, we had a final sample for analysis of 10,992 survey responses.

### 3.2. Comparing Datasets

The final sample that we used for analysis contains 10,992 records. Of these, 10,232 were from the first survey implementation, advertised in the Galaxy Zoo newsletter and with the incentive of beta access to Galaxy Zoo 2, and 760 were from the second survey implementation, advertised in the Galaxy Zoo forum and without the incentive.

In addition to these two phases of survey implementation, we had two variations of the survey instrument, asking demographic questions at the beginning or at the end (section 2.2). Thus, each response set is in one of four conditions. Are there any statistically significant differences in response patterns among these four conditions? If not, all conditions can be combined for analysis; if so, then the conditions should be analyzed separately for any variables that show significantly different response patterns.

For our statistical tests, we used a Kolmogorv-Smirnov (K-S) test for interval and scale variables, and a chi-squared test for nominal variables. These tests were chosen because they do not make any assumptions about the underlying structure of distributions, which is important since our data could be highly non-normal.

No significant differences (at the $p<0.05$ level) were found in any variable between responses to the two versions of the survey instrument.

Next, we used the same statistical tests to compare the two survey implementations, with their different advertising mechanisms and incentives. We found no statistically significant differences between the two samples for any demographic variables, or for the forced choice primary motivation data. Significant differences were found for three of the twelve Likert Scale motivation variables - *Community*, *Fun*, and *Astronomy* (see Table 1 and section 5.2). We therefore performed the analysis of these three Likert Scale variables separately for the two implementation phases, and combined data for the two implementations in the analysis of all other variables.

## 4. RESULTS: DEMOGRAPHICS

Demographic variables can offer a picture of the Galaxy Zoo population, offering insights into who participates in the Galaxy Zoo citizen science project. In our survey, we asked about respondents' gender, age, country of residence, and level of education.

Statistical analyses were performed with SPSS, except for chi-squared tests for statistical significance of demographic variables, and other exceptions stated below. All reported results should be considered in light of the fairly low (6.3%) response rate of the survey.

### 4.1. *Gender and Age*

Gender and age are key characteristics that help us determine our audience. Table 2 shows the gender distribution of responses (n=10,708). More than 80% of respondents to the gender question self-reported as male. An additional 103 respondents reported “prefer not to answer.”

| **Gender** | **Frequency** | **Percent** |
|---|---|---|
| Male | 8,705 | 82.1% |
| Female | 1,900 | 17.9% |

Table 2. Distribution of genders in the sample (n=10,605).

Figure 2 shows a histogram of self-reported age (n=10,952) from age 18 to 80; an additional 40 respondents reported their age as “over 80.” Excluding these “over 80” responses, the mean age is 43.02 years, with a standard deviation of 14.58 years.

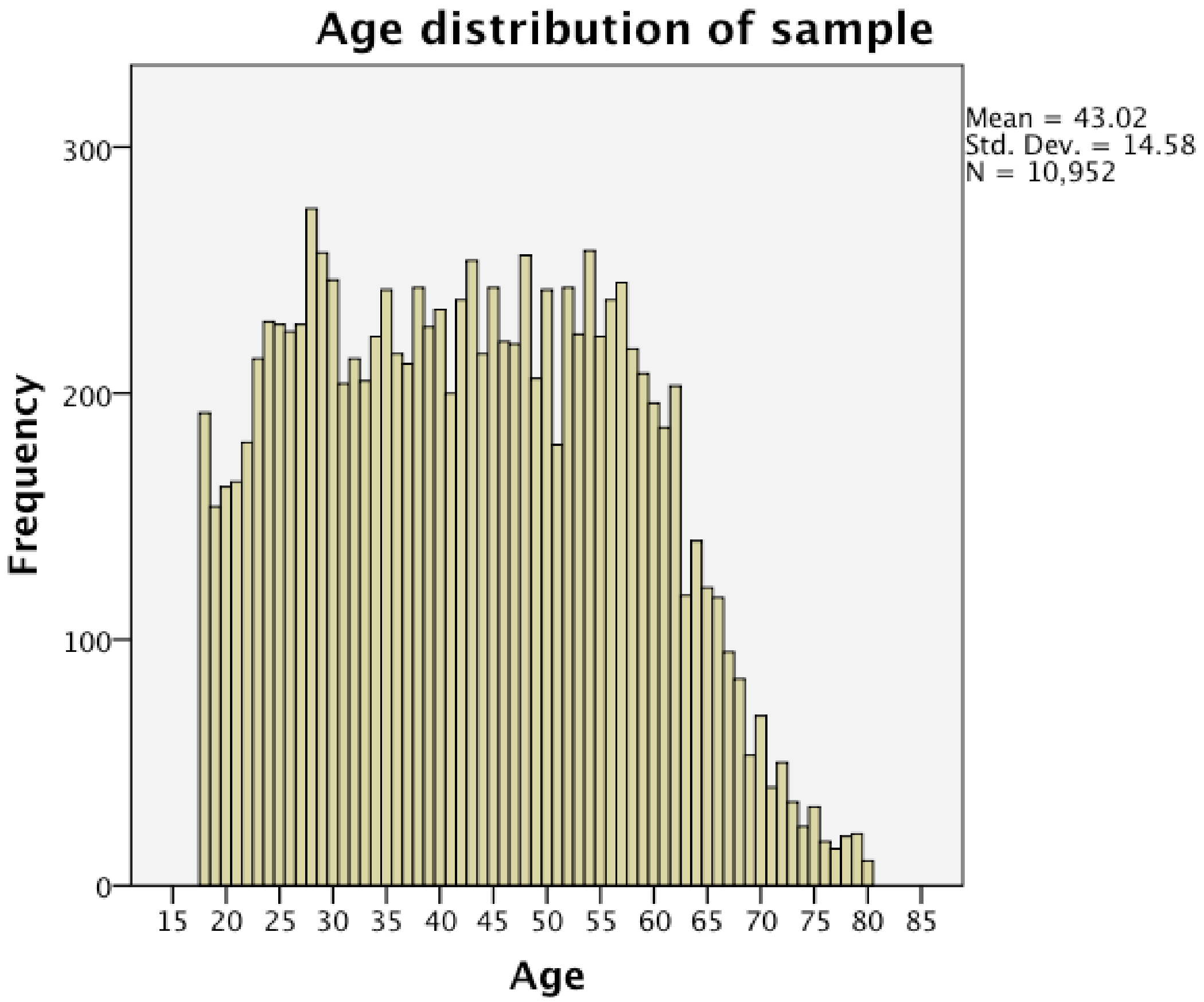

Figure 2. Histogram of age responses in the sample (n=10,952). Ages range from 18 to 80 (responses of “over 80” are excluded). The trend of age responses is fairly flat up to about age 60, then falls off quickly for older ages. The mean age is 43.02, and the standard deviation is 14.58 years.

*4.1.1. Gender and age: interactions*

What is the relationship between the age and gender of survey responses? Figure 3a shows the age distributions of men and women separately (n=10,569), again excluding responses of "over 80" for age and "prefer not to answer" for gender. Figure 3a shows that survey responses from men outnumber those from women at all ages, and that the number of responses for both genders decreases markedly for ages above 60.

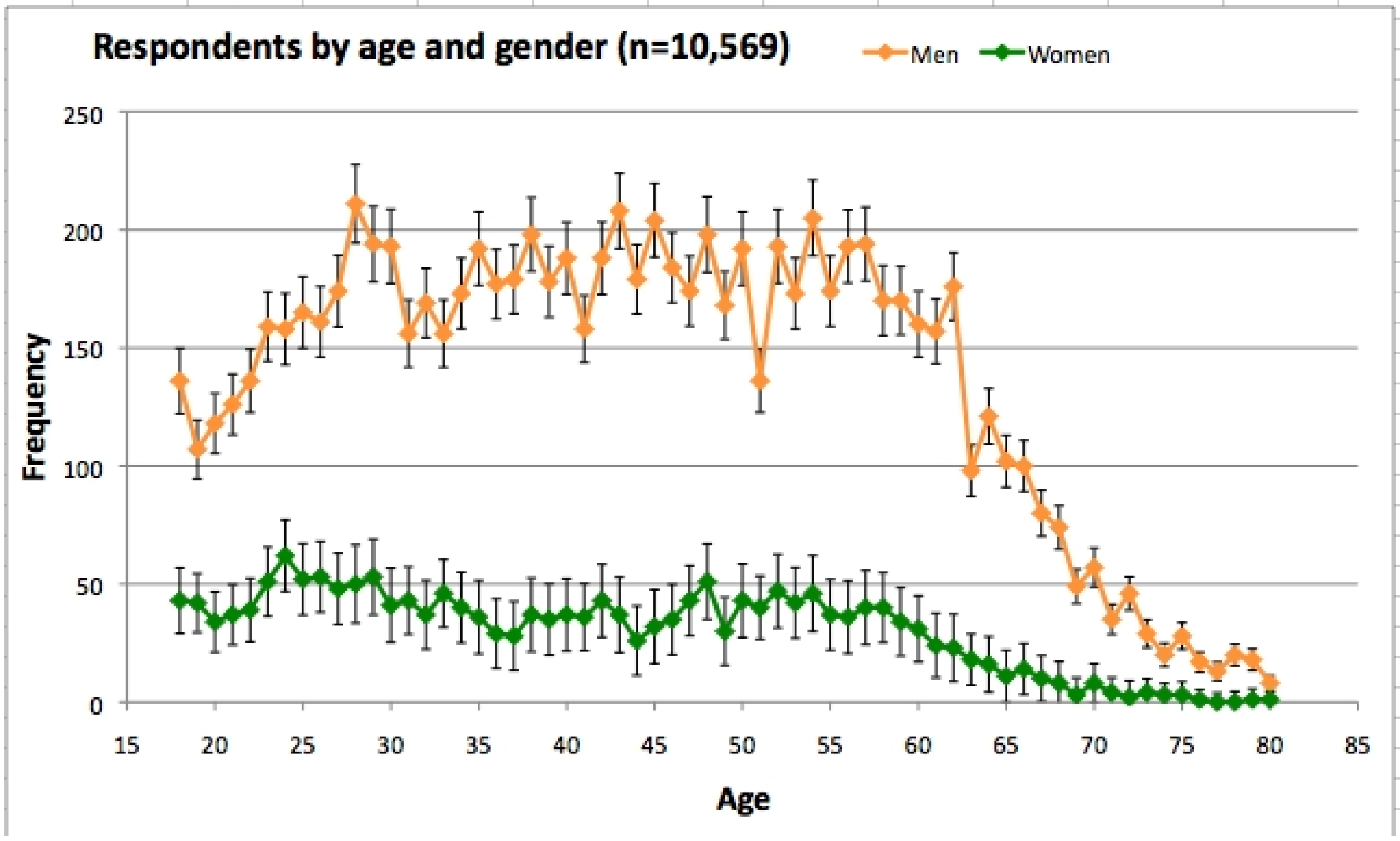

Figure 3a. Gender of survey respondents at each age, from 18 to 80 (n=10,569). The orange circles show the number of of males at each age, and the green circles show number of females at each age. The error bars are the square root of the number of total respondents at each age. The number of responses for both genders decreases markedly with increasing age. Men outnumber women at all ages.

Figure 3b shows the percentage of responses from men and women for each age, from 18 to 80. Men outnumber women at all ages from 18 to 80; but the percentage of men increases with increasing age, from 70% at age 19 to more than 90% at age 80. In other words, at older ages, a larger percentage of respondents self-report as men.

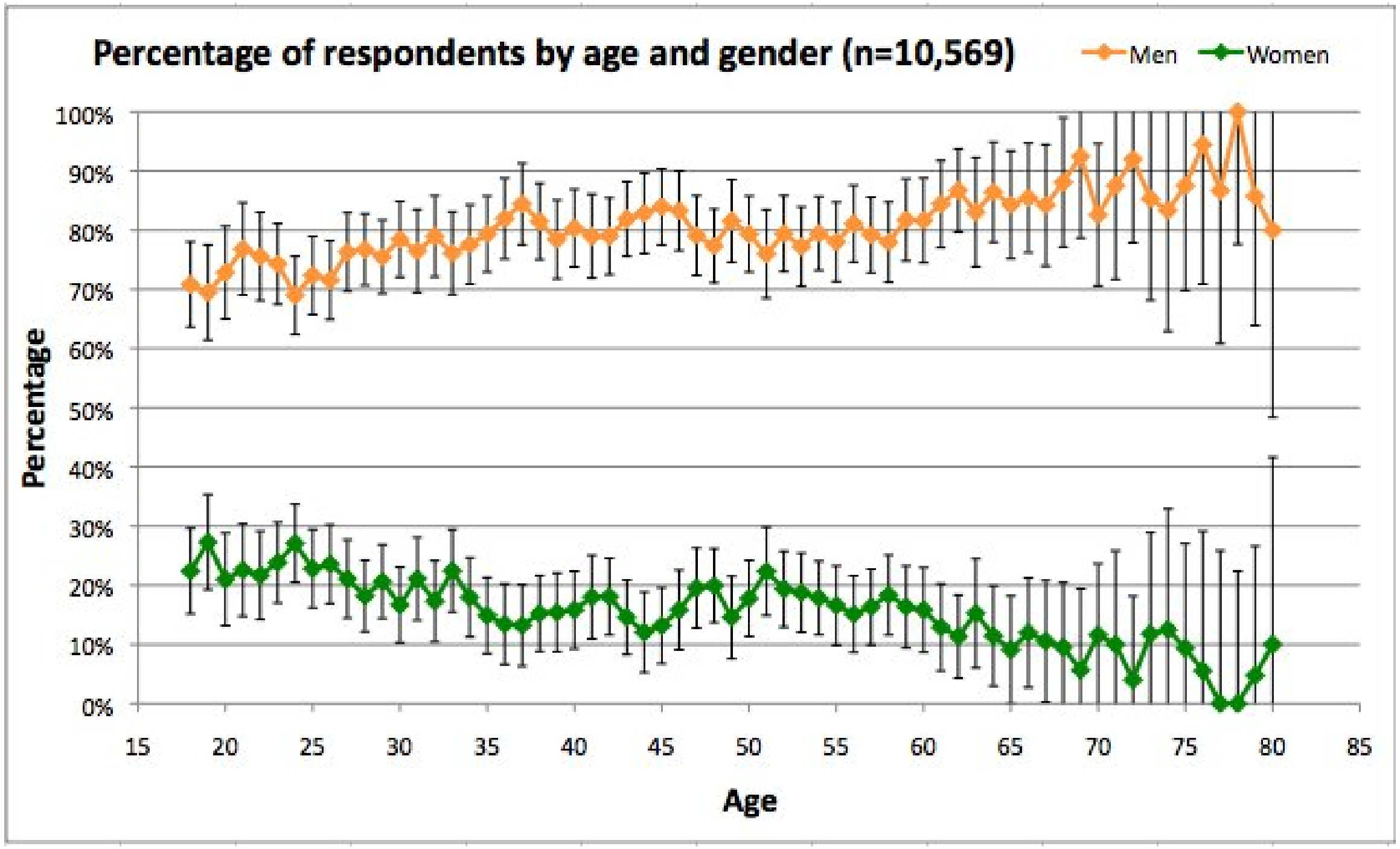

Figure 3b. Percentage of survey-takers self-reporting as male and female at each self-reported age, from 18 to 80 (n=10,569). The error bars are $\frac{\sqrt{n}}{n}$, where n is the total number of respondents at each age. The orange circles show percentage of males and the green circles show percentage of females. The ratio between males and females steadily increases with increasing age.

4.1.2. *Gender and age: comparison to the U.S. online population*

How does the Galaxy Zoo population compare to the overall population of Internet users? To answer this question, we compare our sample's gender and age to that of U.S. Internet users. Although Galaxy Zoo is an international project, more than one third of survey respondents were from the U.S. (section 4.2), and there is no reason to assume that Galaxy Zoo volunteers in the U.S. are systematically different from those in other countries; we therefore compared these volunteers to data about the U.S. population, which is easily available online.

To make this comparison, we selected only responses that reported both age and gender, and reported their country of residence as the U.S. (n=3,756). We compared these responses to what we would find in a representative sample of U.S. adult Internet users, calculated from census data (U.S. Census Bureau 2009) and percentages of adults who use the Internet (Pew Research Center 2010). Estimated errors in these datasets are less than 1% in the Census data and 2.3-2.6% in the Internet use data; considering maximum possible error values when calculating our expected counts in each bin did not change any of our chi-squared significance tests.

Compared to the population of all U.S. Internet users, Galaxy Zoo volunteers in our survey are more than 80% male ($p < 0.001$ preferentially male, according to a chi-squared test).

Next, we compared our responses for gender and age to what we would find in a sample of the same 80/20 gender distribution, but with an age distribution representative of the U.S. online population. A chi-squared test again confirms that the age distribution of our sample does indeed differ from that expected from U.S. Internet users ($p < 0.001$).

Knowing that age distribution differs from expected, in what way does it differ? What groups are over- or under-represented in our sample? Table 3 shows the distribution of survey responses by age bin. Residuals are quoted both as absolute excess/deficit vs. expected and as standardized residuals (Crewson 2006), which provide a measure of which categories are the most important contributors to the observed chi-squared significance. A statistical rule of thumb is that categories whose counts have standardized residuals greater than about 2 are significantly different from their expected values (Crewson 2006); categories whose standardized residuals are greater than 2 are bold in table 3.

| **Age bin** | **Men** | **Excess / deficit vs. expected** | **Women** | **Excess / deficit vs. expected** |
|---|---:|---|---:|---|
| 18-19 | **63** | **-65 (-5.74)** | 29 | -8 (-1.30) |
| 20-24 | **214** | **-87 (-5.01)** | 84 | 1 (0.06) |
| 25-29 | **259** | **-41 (-2.35)** | 109 | 27 (3.04) |
| 30-34 | 265 | -9 (-0.54) | 64 | -13 (-1.47) |
| 35-39 | 274 | -6 (-0.38) | 67 | -1 (-0.13) |
| 40-44 | 303 | 15 (0.88) | 65 | -9 (-1.07) |
| 45-49 | 325 | 17 (0.97) | 97 | 13 (1.46) |
| 50-54 | **364** | **73 (4.30)** | 97 | 17 (1.92) |
| 55-59 | **370** | **122 (7.77)** | 83 | 15 (1.85) |
| 60-64 | **268** | **59 (4.09)** | 55 | 4 (0.54) |
| 65-69 | 168 | 18 (1.44) | **20** | **-16 (-2.67)** |
| 70-74 | **69** | **-42 (-4.01)** | **9** | **-14 (-2.93)** |
| 75-79 | **32** | **-54 (-5.84)** | **3** | **-17 (-3.74)** |

Table 3. The observed age distributions of U.S. men (n=2,974) and women (n=782) in our survey data, compared to what would be expected from a representative sample of the U.S. online population with the same gender ratio as our sample (82% men overall, as in Table 2). The table shows each age bin, the observed number of men and women, and the residuals (positive = excess over expected; negative = deficit under expected). The numbers in parentheses are standardized residuals (Crewson 2006). Bolded values indicate standardized residuals larger than 2. The bold standardized residuals show a deficit of older volunteers in both genders.

Among men, results also show a deficit of younger volunteers and a clear excess of volunteers age 50 to 65.

Table 3 summarizes the comparison between the age and gender of volunteers in our sample with that of the U.S. online population. The table shows that there is a significant excess vs. expected in our data comes from men between the ages of 50 and 65, and significant deficits in responses from men under 30, as well as from all volunteers over age 70.

## 4.2. Geographic distribution

Galaxy Zoo volunteers who filled out the survey reported 118 different countries or territories of residence (n=10,632). The distribution of countries of residence is extremely uneven, with two countries accounting for more than 65% of responses and 25 countries accounting for nearly 95% of responses (Table 4).

| **Rank** | **Country** | **Frequency** | **Percentage** |
|---:|---|---:|---:|
| 1 | United States | 3,834 | 36.1% |
| 2 | United Kingdom | 3,168 | 29.8% |
| 3 | Canada | 439 | 4.1% |
| 4 | Germany | 412 | 3.9% |
| 5 | Australia | 394 | 3.7% |
| 6 | France | 245 | 2.3% |
| 7 | Poland | 233 | 2.2% |
| 8 | Netherlands | 195 | 1.8% |
| 9 | Italy | 135 | 1.3% |
| 10 | Ireland | 114 | 1.1% |
| 11 | Brazil | 99 | 0.9% |
| 12 | Finland | 90 | 0.8% |
| 13 | China | 80 | 0.8% |
| 14 | India | 77 | 0.7% |
| 15 | Spain | 73 | 0.7% |
| 16 | Denmark | 64 | 0.6% |
| 17 | Belgium | 61 | 0.6% |
| 18 | Sweden | 55 | 0.5% |
| 19 | Austria | 53 | 0.5% |
| 20 | South Africa | 49 | 0.5% |
| 21 | New Zealand | 47 | 0.4% |
| 21 | Greece | 47 | 0.4% |
| 23 | Hungary | 43 | 0.4% |
| 24 | Russia | 40 | 0.4% |
| 25 | Switzerland | 38 | 0.4% |

Table 4. The top 25 countries of residence represented in the sample. Percentages calculated exclude missing values and "prefer not to answer" responses. Together, responses from these 25 countries make up 95% of the entire sample (n=10,632).

This geographic distribution is shown on a world map in figure 4.

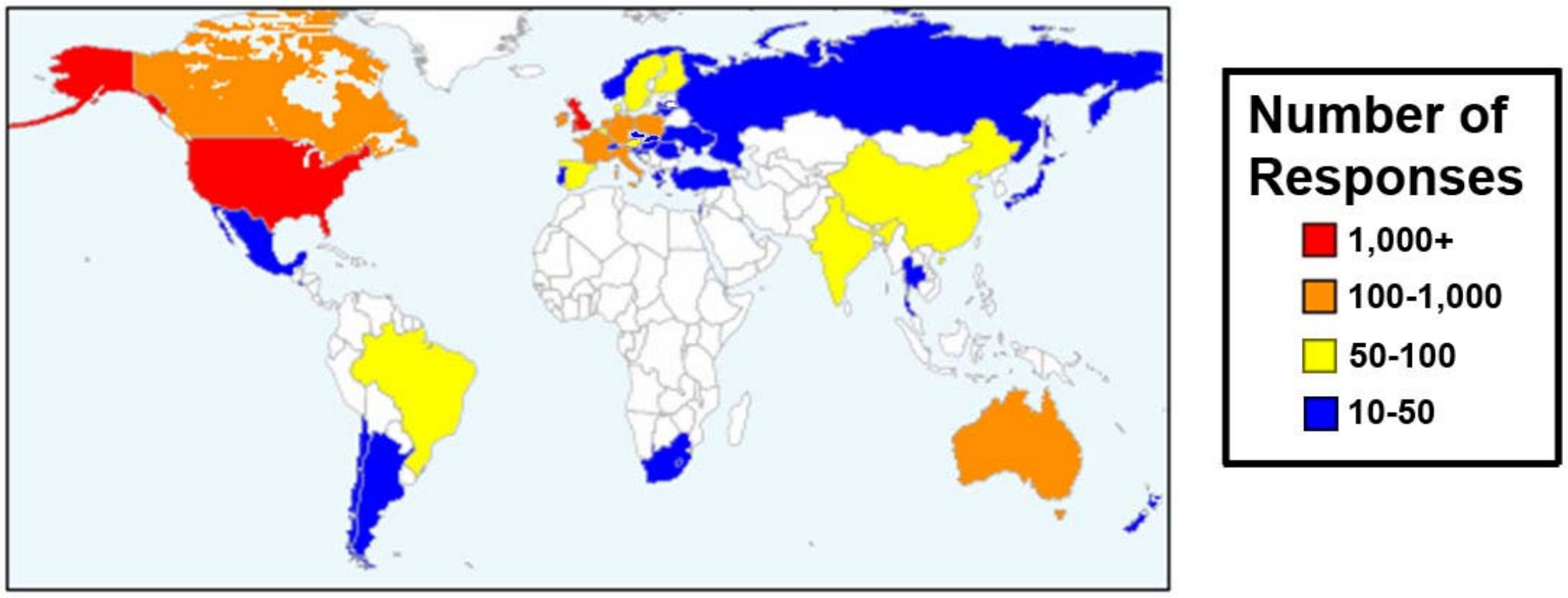

Figure 4. A world map showing all countries with at least 10 survey respondents reporting as their country of residence. The map is color-coded by number of respondents. Red represents more than 1,000 respondents; orange is 100 to 1,000; yellow is 50 to 100; and blue is 10 to 50. The map shows that the reach of Galaxy Zoo extends beyond just the U.S. and U.K.

Figure 4 shows a concentration of responses from richer countries; 89.6% of responses come from countries with per capita GDP greater than $30,000 per year (per capita GDP estimates are purchasing power parity [PPP] estimates from the 2008 CIA World Factbook [Central Intelligence Agency 2008]). These data do not, of course, reflect the income levels of the volunteers themselves. Rather, they provide a measure of the project's global reach; Galaxy Zoo has had some reach into developing nations, but only a limited amount.

## 4.3. Education

### *4.3.1. Education levels by educational system*

An important audience measure for any project is the education level of its audience. The wide geographic distribution of volunteers (section 4.2), combined with the variety of education systems in the world, made recording educational levels challenging. We chose to offer respondents the choice of reporting their education level in either U.S. or U.K. systems. Results for U.S.-like education systems are shown in table 5a; results for U.K.-style education systems are shown in table 5b. The Galaxy Zoo volunteers spanned a large range of educational levels, with the majority having college degrees.

| **Education Level (U.S. and U.S.-like systems)** | **Frequency** | **Percent** |
|---|---|---|
| Student: In High School | 166 | 2.6% |
| Student: Technical / Vocational School | 126 | 2.0% |
| Student: Undergraduate | 577 | 9.0% |
| Student: Postgraduate Program | 67 | 1.1% |
| Student: Graduate School | 170 | 2.7% |
| Student: Professional School (Law, Medical, …) | 68 | 1.1% |
| No High School Degree/ Not a Student | 37 | 0.6% |
| High School diploma or GED | 268 | 4.2% |
| Some college / Not a Student | 583 | 9.1% |
| Technical/vocational certification | 333 | 5.2% |
| Associates Degree | 278 | 4.4% |
| Bachelors degree (BA, BS, …) | 1,748 | 27.4% |
| Postgraduate Professional Certification | 201 | 3.2% |
| Master's degree (MA, MS, …) | 1,121 | 17.6% |
| Doctoral degree (Ph.D., D.Phil., …) | 453 | 7.1% |
| Professional degree (M.D., J.D., …) | 184 | 2.9% |

Table 5a. Frequency of responses to the question, "What is the highest level of education you have completed? (U.S. and U.S.-like school systems)," excluding "prefer not to answer" (n=6,380).

| **Education Level (U.K. and U.K.-like systems)** | **Frequency** | **Percent** |
|---|---|---|
| Student: In Secondary School | 206 | 4.3% |
| Student: In degree program | 330 | 6.9% |
| Student: Postgraduate Study | 128 | 2.7% |
| Student: Graduate School | 94 | 2.0% |
| Left school at 16 (GCSE / O levels) | 393 | 8.3% |
| Left school at 18 (e.g. A levels, highers, …) | 449 | 8.9% |
| Some University / Not a Student | 484 | 9.6% |
| First degree | 1339 | 26.6% |
| Masters degree | 961 | 19.1% |
| Doctoral degree | 378 | 7.5% |
| Prefer not to answer | 279 | 5.5% |

Table 5b. Frequency of responses to the question, “What is the highest level of education you have completed? (U.K. and U.K.-like school systems), excluding “prefer not to answer” (n=4,766).

*4.3.1. Education levels by educational system*

Table 6 shows the self-reported education levels of U.S.-based Galaxy Zoo volunteers 25 years of age or older, and compares these observed responses to the education levels that would be expected from a the same size sample representative of U.S. Internet users (U.S. Census Bureau 2009, Pew Research Center 2010). (We limited this analysis to volunteers 25 and older to enable direct comparison to the census data).

Table 6 shows the residuals and standardized residuals for each education level used by the census data. A chi-squared test ($p < 0.001$) shows that the education levels of respondents are significantly different from that of the U.S. online population; the standardized residuals for each education level show which levels are responsible for the difference.

| **Education Level of U.S. respondents age 25 or over** | **Observed** | **Excess / deficit vs. expected** |
|---|---|---|
| **No High School Diploma** | **6** | **-195 (-13.7)** |
| **High School Graduate** | **116** | **-625 (-23.1)** |
| **Some College or Associate’s Degree** | **773** | **-138 (-4.6)** |
| **Bachelor’s** | **951** | **333 (13.5)** |
| **Postgrad** | **1,003** | **625 (32.3)** |

Table 6. The observed distributions of education levels of survey respondents from the U.S. age 25 or older (n=2,849), compared to what would be expected from a representative sample of U.S. Internet users. The table shows each age bin, the observed number of men and women, and the residuals (positive = excess over expected, negative = deficit under expected). The numbers in parentheses are standardized residuals. Bolded values indicate standardized residuals larger than 2. Across all education levels, Galaxy Zoo volunteers are more educated than the overall online population.

As in our previous analysis, the census data had reported errors of less than 1%, and the Internet use data had reported errors of 2.3-2.6%. Incorporating these errors into expected counts in each bin did not change the significance of any statistical test.

Survey respondents are educated to a higher level that the average adult U.S. Internet user across all levels of education, with nearly 70% of survey respondents (age 25 or older and residing in the U.S.) reporting at least a bachelor's degree.

## 5. RESULTS: MOTIVATIONS OF GALAXY ZOO VOLUNTEERS

Our primary research question was to understand the motivations of citizen scientists taking part in the Galaxy Zoo project: what are the reasons why people have devoted so much time and energy to this citizen science project? We approached this question in several ways, detailed below.

First, in section 5.1, we explore whether there were other motivations present in our population, beyond the twelve listed in Table 1. Then, to explore how survey respondents valued different motivations relative to one another, we analyze Likert Scale data on the importance of each motivation (section 5.2). Next, to compare groups of volunteers by motivation, we analyze forced choice primary motivation data (section 5.3). Finally, we combine these data sources to offer a fuller picture of volunteers' self-reported motivations for participating in Galaxy Zoo (section 5.4).

### 5.1. Searching for additional motivations

The twelve categories of motivation used in this research (Table 1) were inductively determined from a sample of interviews and forum posts, as detailed in Raddick et al. (2010). Those categories reflect themes from those interviews and posts. Since our survey collected data from a much larger sample than our prior work, it is possible that some volunteers in our survey sample might have motivations that were not present in the interview and forum data that were used to derive the categories. If so, then asking questions using our existing twelve categories would miss some of this potentially important information. Therefore, we wanted to give volunteers taking the survey the opportunity to report motivations other than the twelve listed on the survey instrument.

As described in section 2.1, we offered survey-takers the opportunity to report other motivations in two ways. First, after the Likert Scale items for each motivation but before the forced choice question, we asked, "Can you think of any other reasons that someone might be interested in Galaxy Zoo?". We received 1,703 free responses to this question.

Second, the forced choice question, "Which of the reasons above is MOST important to you for participating in Galaxy Zoo?," included an "Other" option, and a free-response textbox to enter another motivation. There were 189 free responses entered to that question.

Having collected these data, the next step was to analyze the responses to discover what additional motivations for volunteering, if any, they might contain. Which responses indicated the same motivations listed on the survey instrument stated in different words, and which indicated genuinely new motivations?

To answer this question, we analyzed all free responses using the same methods we had used to derive the twelve original categories used on the survey instrument – by having three raters develop a consensus classification scheme for the free responses, then independently apply this scheme to the entire corpus of data. We conducted this analysis separately for the two different free response questions, beginning with the responses entered as "Other" options in the forced choice motivation question.

Analysis of these free responses consisted of two rounds, each using the same three independent raters (the raters were the primary author, an undergraduate education major, and a professor of education). Due to a mistake in sample selection, the original analysis took place using a dataset that included responses from people under 18, as well as potential duplicate responses. After seeking input from the Johns Hopkins Institutional Review Board, we removed these motivations from reported results. All results reported in this paper come only from respondents age 18 and over.

In the first round, each free response was independently classified by the three raters using one or more of the following categories: (a) One or more of the existing twelve motivations, expressed in different words; (b) a response indicating 'all of the above'; (c) non-pertinent; or (d) a genuinely new motivation. If a rater chose to include option (d) in a given free response – in other words, if the rater noticed a motivation that he believed to be separate from the existing twelve – he created a word or phrase to describe that motivation.

The raters then shared their classification schemes and discussed all free responses that any one of them had marked as containing a new motivation. Through this discussion, they came to a consensus on a categorization scheme that reflected their shared understanding of the data. That consensus produced four potential new motivations.

In round two, the raters returned to the same set of free response data, and reclassified each response. In round two, the raters chose only a single motivation that they thought best summarized the response.

Each rater independently coded each response as follows: (a) a single specific motivation from the original twelve used in the survey instrument; (b) all of the above (referring to the 12 existing categories); (c) non-pertinent; or (d) one of the four new consensus categories – each response was coded as one of a total of 18 (=12+1+1+4) categories. All three raters assigned one of these 18 categories to a total of 188 free responses.

All three raters agreed on the same category for 84 free responses (45% of the total n=188). Two out of three raters agreed on the same category for 88 free responses (47%). The pairwise agreements between the first and second raters, between the first and third, and between the second and third are 74%, 53%, and 54% respectively. This agreement results in a Fleiss Kappa (Fleiss 1971) interrater reliability of 0.661.

Any responses coded by at least two of the three raters in round two as being in one of the four new consensus categories was judged to reflect a truly new category of motivation. One of the four new categories failed to reach this level of consensus, leaving three.

The final outcome of this two-round coding scheme was three new categories of motivation, together spanning 15 free responses: “Exploration” (motivations associated with exploring space or the universe), “Name a Galaxy” (a desire to have a galaxy or other object named after the respondent), and “Spiritual/Religious.”

Responses judged by the consensus of the raters as new motivation categories thus account for only 15 of the nearly 11,000 responses to the question about a volunteer’s most important motivation. We therefore conclude that our twelve categories are in most cases sufficient to represent the range of motivations that our population would report as primary.

Although this analysis increased our confidence in proceeding with our analysis using our existing twelve categories, we further tested our categories’ completeness by analyzing our other set of free responses motivation data: answers to the question, “Can you think of any other reasons that someone might be interested in Galaxy Zoo?”. The same three raters followed the same two-round procedure to code responses to this question.

Again, the raters independently classified each motivation in round one, naming any new motivations they encountered; then, the raters came to a consensus. The consensus motivation categories found in the first round of analysis of this question included the same three new categories discussed previously, plus several additional categories.

In the second round, as before, the three raters reclassified all free responses using the existing twelve categories and the new consensus categories that came out of the first round, as well as “all of the above” and “prefer not to answer.”

There were 1,645 free responses to which all three raters assigned a category. The independent classifications of the three raters agreed for 456 free responses (28% of the total). Two out of three raters agreed for 929 responses (56%). No agreement was reached for 260 responses (16%). The pairwise agreements between the first and second raters, between the first and third, and between the second and third are 61%, 34%, and 44% respectively. This agreement results in a Fleiss Kappa (Fleiss 1971) interrater reliability of 0.530.

Each new consensus category which had at least one response for which at least two of the three independent raters agreed was included in the final list of potential new motivations. Table 7 shows the twelve new categories that resulted from the raters’ consensus. These twelve potential new motivations account for 316 free responses (19% of the total of 1,645 free responses, and only 3% of the total of 10,992 survey responses). This response level indicates that these additional motivations should perhaps be included in future survey instruments, but that they are unlikely to affect the conclusions described here.

| **New motivation category** | **Explanation** |
|---|---|
| Competition | Desire to have more classifications than others |
| Curiosity | Curiosity about the universe |
| Discover something new | Searching for something that scientists have not seen before |
| **Exploration** | Motivations associated with exploring space or the |
| | universe. |
| Fame | Seeking fame for discoveries |
| Future | Contributing to the future of humanity |
| **Name a Galaxy** | Desire to have a galaxy or other object named after self |
| Novelty | Desire to find a new experience |
| Relaxing | Relaxing or meditative |
| School Assignment | Completing a K-12 or college assignment |
| Self-improvement | Respondent wants to improve themselves |
| **Spiritual/religious** | Spiritual or religious motivations |

Table 7. New motivation categories identified from the question, "Can you think of any other reasons someone might be interested in Galaxy Zoo?". **New Motivation Category** is the consensus name used by the raters, and **Explanation** describes the motivation. Motivations in bold also appeared in the analysis of free responses to the "Other" responses of the primary motivation question. The level at which these additional motivations occur suggests that they are unlikely to be a major source of error in our analysis.

The category "Discover Something New" indicates motivations that discuss a desire to make an original discovery with Galaxy Zoo data, like the Galaxy Zoo volunteer who discovered Hanny's Voorwerp (Lintott et al. 2009). The potential new motivation of "Discover Something New" contrasts with the existing motivation category *Discovery*. The stem given for *Discovery* on the survey instrument read, "I can look at galaxies that few people have ever seen before" – focused on seeing an object that is new to the volunteer, but not necessarily new or unexpected to professional scientists. Nevertheless, these two motivation categories express very similar constructs, and should be carefully separated in future work.

Our analysis of these free response motivations shows some additional, less common, motivations exist in the Galaxy Zoo volunteer population. It also provides some insight to shades of meaning of motivations such as *Discovery* and "Discover Something New." However, our analysis uncovered potential new motivations only at the few percent level. We thus conclude that our original twelve motivations are largely sufficient to cover motivations in our study sample. We therefore proceed with analysis of our dataset using the twelve motivations from Table 1.

## 5.2. Likert Scale results

### *5.2.1. Overall results*

Figures 5a-c show histograms of responses to the question "how motivating are these factors to you?," for Likert Scale responses to each of the 12 motivation categories. Each response ranges from 1 (anchored as "not at all motivating") to 7 ("very motivating"). As described in section 2.2, we had two survey implementations: one with a survey completion incentive that was advertised through a Galaxy Zoo newsletter, and one without an incentive that was advertised through the Galaxy Zoo forum. For nine of the twelve motivations, statistical tests showed that data could be combined between the two implementations (section 3.2); histograms of Likert Scale responses for those motivations are shown in Figure 5a. Figures 5b-c show results for the other three motivations, separated by survey implementation. For these three motivations, responses trended higher (more motivating) in the implementation with a sample of forum users without an incentive for completing the survey.

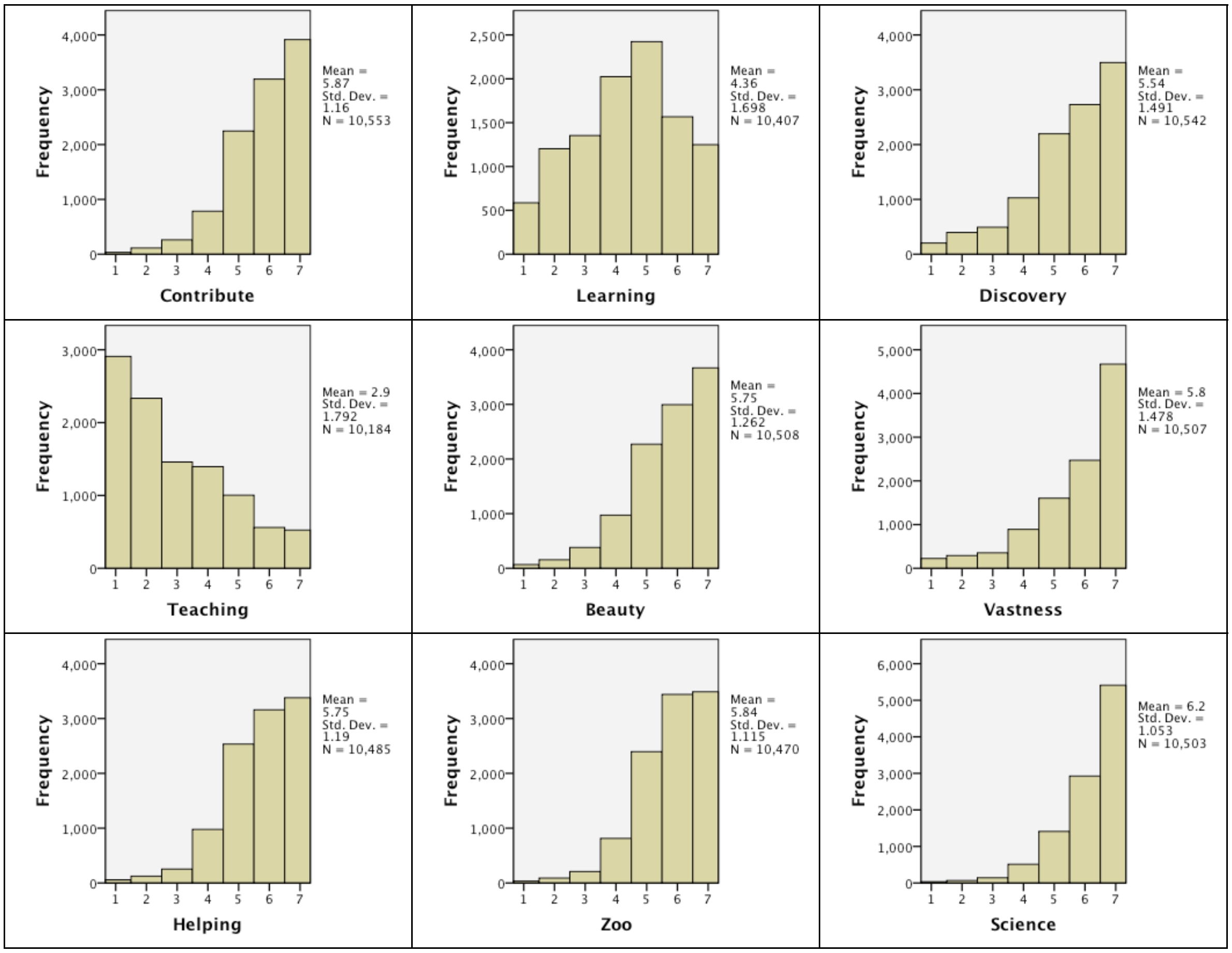

Figure 5a. Histograms showing responses to Likert Scale questions for the nine motivations in which responses could be combined for the two survey implementations. The histograms show that respondents identified several motivations as being important.

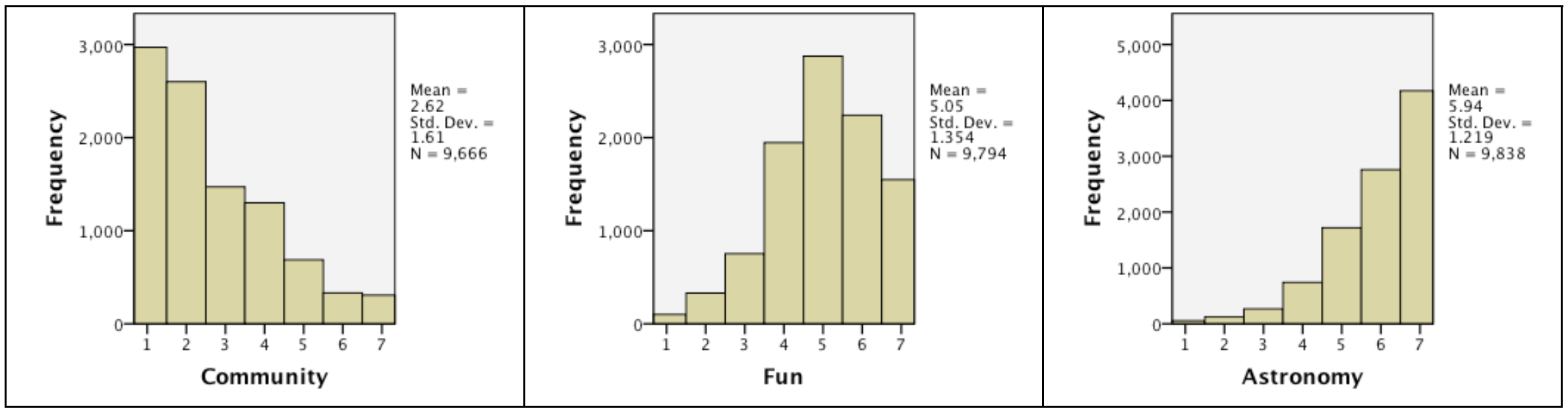

Figure 5b. Histograms showing responses to Likert Scale questions for three motivations in the sample resulting from the implementation of the survey with the overall volunteer population, and with the incentive of Galaxy Zoo 2 beta access.

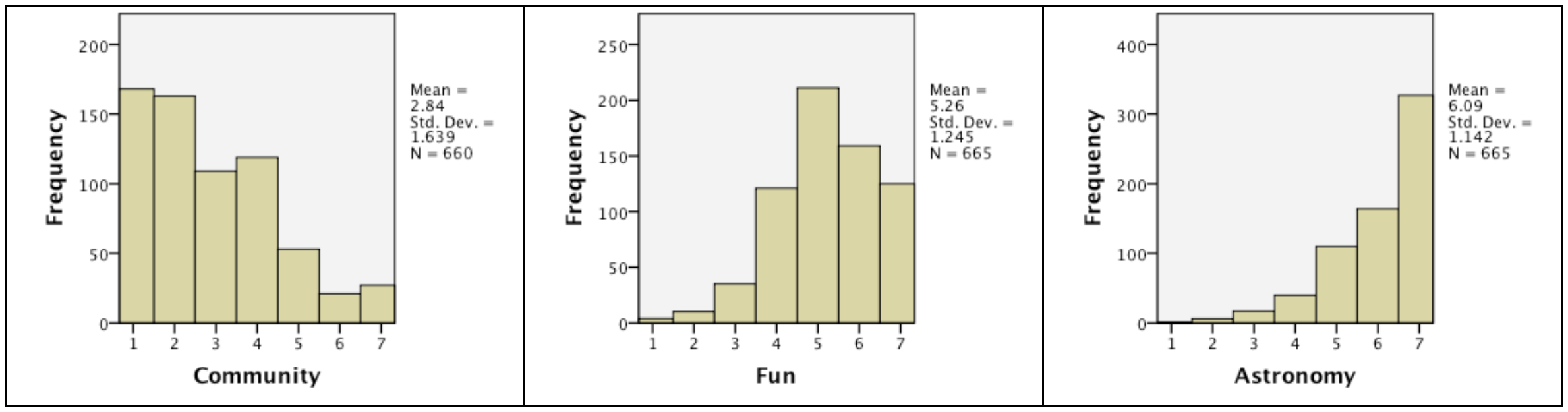

Figure 5c. Histograms showing responses to Likert Scale questions for three motivations in the sample resulting from the implementation of the survey with the forum user population, and without the incentive.

Table 8 shows the correlation matrix among the twelve motivations (the correlations quoted are Spearman rho correlations, appropriate for ordinal variables like Likert Scales). The motivations display low intercorrelations (rho < 0.7), which suggest that the motivation categories measure constructs that are generally independent of one another.

| | Contribute (A) | Learning (B) | Discovery (C) | Community (D) | Teaching (E) | Beauty (F) | Fun (G) | Vastness (H) | Help (I) | Zoo (J) | Astronomy (K) | Science (L) |
|---|---|---|---|---|---|---|---|---|---|---|---|---|
| **A** | 1.000 | 0.259 | 0.336 | 0.145 | 0.141 | 0.226 | 0.280 | 0.261 | 0.450 | 0.486 | 0.358 | 0.421 |
| **B** | | 1.000 | 0.326 | 0.462 | 0.396 | 0.308 | 0.348 | 0.339 | 0.310 | 0.358 | 0.339 | 0.238 |
| **C** | | | 1.000 | 0.243 | 0.212 | 0.513 | 0.408 | 0.460 | 0.341 | 0.373 | 0.282 | 0.242 |
| **D** | | | | 1.000 | 0.536 | 0.201 | 0.288 | 0.217 | 0.226 | 0.241 | 0.263 | 0.161 |
| **E** | | | | | 1.000 | 0.222 | 0.287 | 0.217 | 0.218 | 0.241 | 0.253 | 0.172 |
| **F** | | | | | | 1.000 | 0.522 | 0.522 | 0.377 | 0.367 | 0.298 | 0.247 |
| **G** | | | | | | | 1.000 | 0.413 | 0.454 | 0.450 | 0.322 | 0.268 |
| **H** | | | | | | | | 1.000 | 0.471 | 0.416 | 0.311 | 0.271 |
| **I** | | | | | | | | | 1.000 | 0.653 | 0.379 | 0.369 |
| **J** | | | | | | | | | | 1.000 | 0.515 | 0.468 |
| **K** | | | | | | | | | | | 1.000 | 0.631 |
| **L** | | | | | | | | | | | | 1.000 |

Table 8. Correlations (Spearman rho) between overall Likert scale scores for each of the twelve motivation categories. Correlations between motivations are present but are all less than 0.7, suggesting that motivations measure related but separate constructs.

These results shown in Figure 5a-c suggest two things. First, Galaxy Zoo volunteers have many reasons for participating in the project. Second, while volunteers perceive some motivations as about equally important (e.g. *Contribute* and *Beauty*), the motivations of *Community*, *Teaching*, and *Learning* are clearly less important to volunteers. This does not mean that volunteers are not learning, or that they are not valuing the opportunity to learn from their participation – only that learning is not a major *reason* for their participation in Galaxy Zoo.

*5.2.2. Likert scale responses by demographics*

To explore whether the motivations of demographic subgroups are different, we found the Likert Scale responses for various subgroups. To enable easier comparison between responses, we standardized the scores for each motivation category into Z-scores (a renormalization of volunteers' Likert Scale responses such that the resulting distribution has a mean of zero and a standard deviation of one); this procedure allows us to compare measures of central tendency for the Likert Scale results. Results are shown in tables 9a-c, disaggregated by gender.

Table 9a shows results for motivation categories in which the two survey samples could be combined. Tables 9b-c show the categories that could not be combined. Each table shows data for men in the first three columns, women in the middle three columns, and combined data (the difference in means and a test of statistical significance) in the last two columns. A negative mean Z-score indicates that a group rated that motivation lower than the mean of the entire sample; a positive mean Z-score means that the group rated that motivation higher than the mean of the entire sample. In the last two columns (men-women), negative values indicate that women rated a motivation more highly; positive values indicate that men rated it more highly.

| | MEN | | | WOMEN | | | MEN-WOMEN | |
|---|---|---|---|---|---|---|---|---|
| Motivation Z-score | N | Mean | SD | N | Mean | SD | $M_{men}$ - $M_{women}$ | Statistically significant? |
| *Contribute* | 8522 | -0.02 | 1.00 | 1874 | 0.11 | 0.99 | -0.14 | YES; p < 0.01 |
| *Learning* | 8395 | -0.03 | 0.99 | 1859 | 0.13 | 1.02 | -0.16 | YES; p < 0.01 |
| *Discovery* | 8512 | -0.04 | 1.01 | 1875 | 0.20 | 0.92 | -0.24 | YES; p < 0.01 |
| *Teaching* | 8215 | -0.01 | 0.99 | 1822 | 0.06 | 1.06 | -0.08 | YES; p=0.02 |
| *Beauty* | 8481 | -0.06 | 1.02 | 1873 | 0.30 | 0.85 | -0.36 | YES; p < 0.01 |
| *Vastness* | 8488 | -0.04 | 1.02 | 1868 | 0.19 | 0.88 | -0.23 | YES; p < 0.01 |
| *Help* | 8467 | -0.04 | 1.00 | 1867 | 0.19 | 0.96 | -0.23 | YES; p < 0.01 |
| *Zoo* | 8457 | -0.04 | 1.00 | 1863 | 0.19 | 0.95 | -0.23 | YES; p < 0.01 |
| *Science* | 8476 | 0.04 | 0.95 | 1873 | -0.15 | 1.16 | 0.19 | YES; p < 0.01 |

Table 9a. Motivation z-scores for men and women. Women rank most motivations significantly more highly than men, which may be a result of acquiescence bias (Hodson 1989), but rank *Beauty* even more highly than other motivations. On the other hand, men rank *Science* significantly more highly than women.

| | MEN | | | WOMEN | | | MEN-WOMEN | |
|---|---|---|---|---|---|---|---|---|
| Motivation z-score | N | Mean | SD | N | Mean | SD | $M_{men}$ - $M_{women}$ | Statistically significant? |
| *Community* | 7808 | 0.00 | 1.00 | 1720 | -0.05 | 1.01 | 0.05 | YES; p < 0.01 |
| *Fun* | 7907 | -0.09 | 1.00 | 1748 | 0.36 | 0.91 | -0.45 | YES; p < 0.01 |
| *Astronomy* | 7955 | 0.01 | 0.98 | 1745 | -0.10 | 1.09 | 0.12 | YES; p < 0.01 |

Table 9b. Likert-scale z-scores for men and women from the sample of all Galaxy Zoo volunteers, with the incentive of beta access to Galaxy Zoo 2.

| | MEN | | | WOMEN | | | WOMEN-MEN | |
|---|---|---|---|---|---|---|---|---|
| Motivation z-score | N | Mean | SD | N | Mean | SD | $M_{women}$ - $M_{men}$ | Statistically significant? |
| *Community* | 534 | 0.13 | 1.00 | 115 | 0.10 | 1.13 | 0.03 | NO |
| *Fun* | 535 | 0.10 | 0.92 | 118 | 0.39 | 0.91 | -0.29 | YES; p = 0.02 |
| *Astronomy* | 536 | 0.13 | 0.92 | 117 | 0.12 | 0.97 | 0.01 | NO |

Table 7c. Likert-scale z-scores for men and women from the sample of forum users, without the incentive of beta access to Galaxy Zoo 2.

Statistically significant differences in mean motivation Z-scores between men and women were found in most cases; exceptions are highlighted in the last column of table 9c. Mean z-scores for women are higher than for men in most categories; this may be a result of acquiescence bias (Hodson 1989), in which women are more likely to rate items higher on Likert Scales simply because doing so is seen as socially desirable.

The motivations with the largest difference between women's and men's ratings are *Beauty* ($M_{men} - M_{women}$ = -0.36 in the combined sample) and *Fun* (-0.45 in the incentivized sample, -0.29 in the nonincentivized sample). The only motivations that men rate higher than women are *Science* ($M_{men} - M_{women}$ = +0.19 in the combined sample) and *Astronomy* (+0.12 in the incentivized sample).

5.3. Motivations of Galaxy Zoo volunteers: Primary Motivations

To allow us to sort volunteers into groups based on motivation, we asked survey-takers to choose the most important reason for their participation in Galaxy Zoo. Results are shown in Table 10, excluding "Prefer not to answer" responses. All users who selected *Other* are listed as *Other* in table 10, even those whose motivations were judged by the three raters to be the original twelve motivations stated in different words (section 5.1).

| **Motivation Category** | **Frequency** | **Percentage** |
|---|---|---|
| *Contribute* | 4,192 | 39.8% |
| *Astronomy* | 1,301 | 12.4% |
| *Discovery* | 1,091 | 10.4% |
| *Beauty* | 939 | 8.9% |
| *Vastness* | 870 | 8.3% |
| *Science* | 718 | 6.8% |
| *Zoo* | 429 | 4.1% |
| *Helping* | 300 | 2.8% |
| *Fun* | 294 | 2.8% |
| *Learning* | 172 | 1.6% |
| *Other* | 135 | 1.3% |
| *Teaching* | 75 | 0.7% |
| *Community* | 16 | 0.2% |

Table 10. Primary motivations for all respondents in the sample (n=10,532). In contrast to the Likert Scale results (figure 5a-c), these results show that ***Contribute*** is clearly the most important motivation to this sample of Galaxy Zoo volunteers.

The most popular response was *Contribute* ('I am excited to contribute to original scientific research'), with nearly 40% of responses; no other motivation category captured more than 13% of responses.

Table 11 shows results for primary motivation disaggregated by gender; *Contribute* is the most frequently cited primary motivation for both men and women.

| | MEN | | WOMEN | | MEN - WOMEN |
|---|---|---|---|---|---|
| **Motivation** | **Frequency** | **Percentage** | **Frequency** | **Percentage** | **Percentage** |
| Contribute | 3,441 | 40.4% | 702 | 37.4% | +3.0% |
| Astronomy | 1,105 | 13.0% | 180 | 9.6% | +3.4% |
| Discovery | 888 | 10.4% | 187 | 10.0% | +0.5% |
| Beauty | 701 | 8.2% | 226 | 12.0% | -3.8% |
| Vastness | 666 | 7.8% | 188 | 10.0% | -2.2% |
| Science | 620 | 7.3% | 92 | 4.9% | +2.4% |
| Zoo | 340 | 4.0% | 87 | 4.6% | -0.6% |
| Helping | 240 | 2.8% | 53 | 2.8% | 0.0% |
| Fun | 199 | 2.3% | 91 | 4.8% | -2.5% |
| Learning | 143 | 1.7% | 26 | 1.4% | +0.3% |
| Other | 106 | 1.2% | 23 | 1.2% | 0.0% |
| Teaching | 53 | 0.6% | 21 | 1.1% | -0.5% |
| Community | 13 | 0.2% | 2 | 0.1% | 0.0% |

Table 11. Primary motivations for men (n=8,515) and women (n=1,878), excluding "prefer not to answer" responses. Both men and women choose ***Contribute*** as their most important motivation. Because these data come from a forced choice question rather than a Likert Scale, they are less vulnerable to acquiescense bias effects than the data in Table 9a-c.

The differences in response patterns between men and women (rightmost column) show a small but statistically significant difference in most important motivation choice (as shown by a chi-squared test with $p<0.01$). The differences result from differences in the following motivations: men are slightly more likely to choose *Contribute* (+3.0%), *Astronomy* (3.4%) and *Science* (2.4%) as primary motivations, while women are more likely to choose *Beauty* (-3.8%), *Vastness* (-2.2%) and *Fun* (-2.5%). Because these results are forced choices rather than based on Likert scale responses, they are less susceptible to the acquiescence bias discussed in the previous section.

However, it is important to emphasize that *Contribute* is by far the most frequently chosen motivation, regardless of gender. *Contribute* is also the most frequently chosen motivation for all ages (18 to "over 80") and all levels of education (in both the U.S. and U.K. systems).

## 5.4. Motivations of Galaxy Zoo volunteers: Comparing Measures

As described in this section and in section 2.1, we chose to ask about volunteer motivation in two ways. To allow volunteers to report all the motivations they felt were important to them, we asked for Likert Scale ratings for each of the twelve motivations in Table 1; these results are reported in section 5.2. To enable us to divide volunteers into motivation-based groups for analysis, we asked volunteers to choose their most important motivation; these results are reported in section 5.3.

Since we have two measures that measure the construct of motivation, we naturally ask how those measures can be combined to gain a fuller understanding of motivation. Figure 6 is a contour plot comparing the results of forced choice primary motivations (section 5.3) with the Likert Scale values for all motivations (section 5.2).

The y-axis of Figure 6, *Primary Motivation*, displays the twelve groups created from responses to the question about most important motivation. The groups are given from top to bottom in order of increasing frequency (Table 10), from *Community* (n=16) to *Contribute* (n=4,192). The x-axis, *Likert Scale Category*, shows the twelve Likert Scale motivation variables, left to right in the opposite order as primary motivation. The colors indicate the mean of each Likert Scale value within each group.

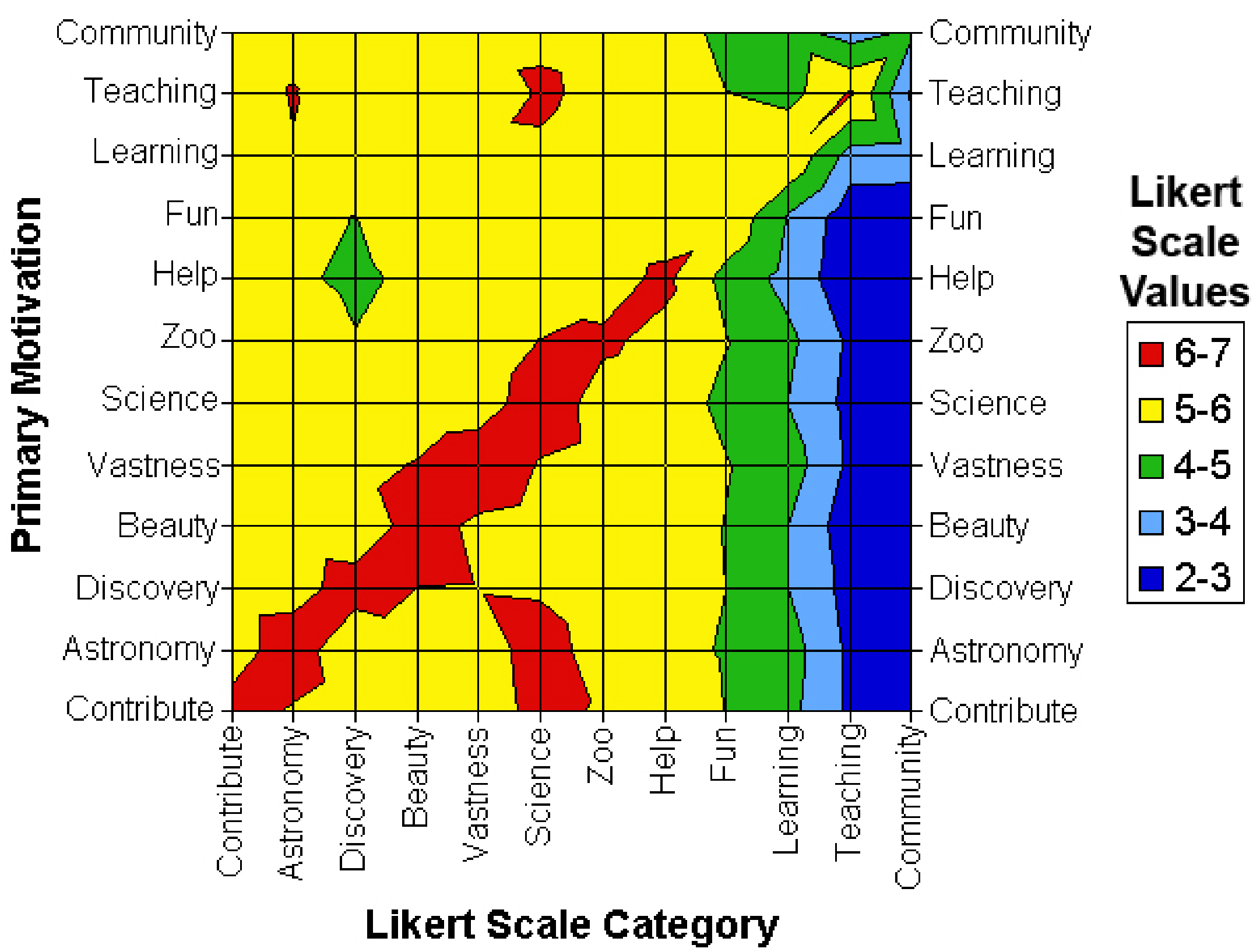

Figure 6. The relationship between primary motivation and Likert Scale data. The y-axis lists the twelve primary motivations, top to bottom, in order of increasing frequency in the entire sample (from least to most common). The x-axis shows the Likert Scale scores for each motivation, left to right, in the same order. Some interesting relationships are discussed in the text.

To read Figure 6 to find the relationship between a primary motivation and the Likert Scale scores of respondents who chose that motivation, read horizontally from one of the twelve primary motivations. Moving along that horizontal line, each grid point gives the mean Likert Scale value for each of the 12 motivations among people who chose that primary motivation. The colors show the mean Likert Scale values: 6-7 is red, 5-6 is yellow, 4-5 is green, 3-4 is light blue, and 2-3 is dark blue. For example, reading along the ninth and eleventh rows of the table to the fifth column, users who chose *Astronomy* as their primary motivation rated *Science* in their Likert Scale motivation responses more highly that users who chose *Beauty*.

As would be expected, the diagonal ridge in Figure 6 (running from bottom left to top right) shows that volunteers who choose a given primary motivation generally rate that same motivation higher in their Likert scale responses as well.

The regions of Figure 6 that are off the diagonal show some interesting relationships between a respondent's chosen primary motivation (along horizontal rows) with their Likert Scale responses for the other eleven categories of motivation (along vertical columns) – motivations that they did not select as primary.

The lists below show some interesting relationships we discovered in this way, indicating the mean (M) and standard deviation (S) of both the subsample of respondents who chose the given primary motivation and the overall sample.

For volunteers who chose primary motivations related to helping with the Galaxy Zoo project (*Contribute* and *Help*):

1) Volunteers who chose *Contribute* as their primary motivation tend to rate *Science* higher (M = 6.33, S = 0.938) than the overall sample of all volunteers (M=6.20, S = 1.053).
2) Volunteers who chose *Help* as their primary motivation tend to rate *Discovery* (M = 4.79, S = 1.580) lower than the overall sample of all volunteers (M = 5.54, S = 1.491).

For volunteers who chose primary motivations related to Galaxy Zoo's scientific content (*Astronomy*), data (*Beauty*), or task (*Fun*):

1) Volunteers who chose *Astronomy* as their primary motivation tend to rate *Science* (M = 6.43, S = 0.880) and *Learning* (M = 4.58, S = 1.609) highly than the overall sample (M = 6.20, S = 1.053; M = 4.36, S = 1.698 respectively).
2) Volunteers who chose *Beauty* as their primary motivation are more likely to rate *Contribute* (M = 5.15, S = 1.282) lower than the overall sample (M = 5.87, S = 1.160).
3) Volunteers who chose *Fun* as their primary motivation are more likely to rate *Contribute* (M = 5.22, S = 1.244) lower than the overall sample (M = 5.87, S = 1.160).

For volunteers who chose primary motivations related to learning and teaching:

1) Volunteers who chose *Learning* as their primary motivation were more likely to rate *Teaching* (M = 3.85, S = 2.072) highly than the overall sample (M = 2.90, S = 1.792).
2) Volunteers who chose *Teaching* as their primary motivation were more likely to rate *Astronomy* (M = 6.15, S = 1.081) highly than the overall sample (M = 6.09, S = 1.142 for the non-incentivized sample of forum users, M = 5.94, S = 1.220 for the incentivized sample of general Galaxy Zoo volunteers).

We discuss possible interpretations of these results, in light of the literature on motivation, in the next section.

## 6. DISCUSSION

This section interprets the results reported in sections 4 and 5. First, we compare the demographics of this sample of Galaxy Zoo users to those from similar online astronomy and/or volunteer collaborative projects. Next, we examine the observed motivations in light of existing motivation literature, and propose several hypotheses about underlying causes of observed motivation responses. These hypotheses can be tested through further surveys and interviews with volunteers in Galaxy Zoo and other citizen science projects.

In interpreting these results, we caution that this study treats Galaxy Zoo volunteers as a single population, even though different volunteers have greatly different ways of being involved in Galaxy Zoo. The results also come from a single point in time; motivations may change with increased time or depth of involvement.

## 6.1 Demographics

We find a significant excess of male participants between age 50 and 65 compared to that of the U.S. online population (Section 4.1.2, Table 3). This excess of men in that age group is also seen in those who participate in amateur astronomy (Price & Paxson 2012).

Responses in our sample are similar to those from a sample of volunteers with Citizen Sky, another online astronomy citizen science project, in which 78% of respondents self-reported as male, with a mean age of 41 and a standard deviation of 16 years (Price 2011). Responses are also broadly similar to responses in a survey sample of listeners to a popular astronomy podcast (Gay et al. 2007), which were 92% male, with an average age of 40. Previous surveys of astronomy podcasts showed responses of 87-91% male (Gay, Price, & Searle 2006). As a non-online comparison sample, subscriptions to Sky & Telescope magazine are 95% male, with a mean age of 51 (New Track Media 2010).

Respondents come from many different countries, with the United States and United Kingdom as the two largest countries (section 4.2 and Table 4).Across all education levels in the U.S., Galaxy Zoo volunteers are significantly more educated than the general online population (section 4.3 and Table 6). That result is not unexpected, given that Galaxy Zoo is an online science project that requires access and comfort with using the Internet, as well as interest and self-efficacy in science.

## 6.2. Motivations

Our analysis of responses to questions about other motivations (section 5.1) confirms that the twelve categories of motivation found in our prior work (Raddick et al. 2010), listed in Table 1, do cover most of the motivations present in the Galaxy Zoo volunteer population. Although some other motivations are undoubtedly present at some level, the existing ones can reasonably explain a very large fraction of volunteer motivations.

In sections 5.2 to 5.4, we present data on the prevalence of these motivations in our sample of survey responses, as well as in various subgroups. To interpret what these results can tell us about motivation to participate in Galaxy Zoo, it is useful to consider the results in light of existing literature about motivation to participate in similar contexts, both online and offline.

The central framework we employ for understanding motivation in this work is expectancy-value theory, which understands a person's motivation to perform an action to be determined by both the result they expect to occur due to that action (expectancy), and the value they attach to achieving that result (Atkinson 1957).

For a freely chosen volunteer activity like Galaxy Zoo, expectancy is determined by the person's self-efficacy – the belief that they can successfully complete a given action (Bandura 1977) – in this case, classifying a galaxy. Value is determined by the benefits they perceive will accrue as a result of that action (Eccles & Wigfield 2002). Some benefits that have been examined in literature on online volunteer activities include altruism and identification with the project's goals (Crowston & Fagnot 2008), interest in the project's content (O'Brien & Toms 2010), and factors related to the project's specific task, such as the variety of skills required and the task's perceived significance (Hackman & Oldham, 1976). Within the wider context of motivation to participate in volunteer activities in general, motivations considered include expressing values, fostering social engagement, and providing opportunities for learning or career advancement (Clary & Snyder 1999; Bang & Ross 2009; Rhoden, Ineson, & Ralston 2009).

The Likert Scale responses for the twelve motivation categories confirm that respondents have many different motivations for participating in Galaxy Zoo, including all the factors described above. However, the general trend in responses (Figure 5a-c) suggests that a greater number of volunteers are motivated by identification with project goals and interest in scientific content, and are less motivated by learning science and participating in a social community. Men are more likely to be motivated by interest in science, while women are more likely to be motivated by desire to see beautiful galaxy images and the fun of classifying galaxies.

But we caution that these results reported as "motivations of Galaxy Zoo users" are motivations measured at a single point in time, from a sample of the entire population of Galaxy Zoo volunteers. In reality, different volunteers engage with citizen science in different ways, for different reasons – and these reasons are likely to change over time. In addition, although learning science in itself does not appear to be a major motivation for Galaxy Zoo participation, these results say nothing about whether and how volunteers learn, or want to learn, science. Further research is needed to understand how motivation and learning happen in citizen science. Members of our research team are currently exploring such questions.

The most exciting result of this research is our conclusion, reported in Table 10, that the most common primary motivation reported by this sample of Galaxy Zoo volunteers is *Contribute*, "I am excited to contribute to original scientific research." *Contribute* was selected by nearly 40% of respondents, at equal levels by both men and women, and was the largest primary motivation for all age bins and education levels.

Although *Contribute* was not the most highly-rated motivation in the Likert Scale results, it was by far the most important primary motivation seen in the forced choice results. Had we asked about motivations using Likert Scale questions only, we would have missed this key discovery. This has major implications for the design of instruments to measure motivations of citizen scientists.

By comparing Likert Scale data and primary motivation data (Figure 6), we discovered interesting relationships between a number of motivations. Volunteers who chose *Contribute* as their primary motivation are more likely to rate *Science* higher; this could indicate some conceptual overlap between and interest in contributing to science and interest in science itself.

Some motivations appear to vary inversely with one another. Volunteers who chose *Beauty* or *Fun* as their primary motivation are more likely to rate *Contribute* lower. This may indicate that volunteers who participate in Galaxy Zoo out of personal enjoyment of the activity – whether it is the fun of classifying or the beauty of the images – are less motivated by making a contribution to science. However, we emphasize that these volunteers still are making a contribution to science through their classifications. Volunteers who chose *Help* as their primary motivation are more likely to rate *Discovery* lower, perhaps indicating that volunteers who view helping Galaxy Zoo as intrinsically worthwhile are less concerned about viewing images that have rarely been seen before.

## 7. FUTURE WORK

Future studies of the motivations of volunteers in Galaxy Zoo and other citizen science projects should continue in at least three directions. First, volunteers' own understandings of each of the twelve motivation categories should be further unpacked. How do the different motivations relate? What does it mean for volunteers to contribute to science? What qualities of the Galaxy Zoo images do volunteers find beautiful? What aspects of the interface are fun for volunteers?

Second, this study considered volunteers as a single population, even though different volunteers have greatly different ways of being involved in Galaxy Zoo. Some classify galaxies only once, some return to classify thousands of galaxies, and some go far beyond the basic task of classifying galaxies into a range of different involvements with the project, such as reading the Galaxy Zoo blog or engaging in deeper scientific research on new astronomical objects in the Galaxy Zoo forum (Cardamone et al. 2009). These deeper involvements may be driven by entirely different motivations from those that motivate volunteering with Galaxy Zoo initially, and volunteer motivations may change with deepening engagement.

Third, other populations of citizen science volunteers should be studied to determine whether and how these results can be generalized beyond Galaxy Zoo. We have already analyzed the results of interviews with volunteers in the Moon Zoo citizen science project. We plan to repeat our study with some of the other citizen science projects that make up Zooniverse, and we strongly encourage other citizen science groups to study the motivations of their volunteer populations as well.

## 8. CONCLUSIONS

We have explored the motivations of citizen scientists who volunteer with the Galaxy Zoo project. Through a survey of about 11,000 existing Galaxy Zoo users, we found that the most frequently cited reason for participating in Galaxy Zoo is a desire to contribute to science. The discovery that contributing to science is a major motivator for citizen science participation is not in itself a surprise, but the fact that nearly 40% of respondents report contributing to science as their primary motivation is stunning.

The implication of this result is that Galaxy Zoo volunteers have a genuine desire to be involved in scientific research. Of all the possible reasons that the Galaxy Zoo website was so successful, the reason chosen as most important by the largest number of survey respondents was not specific to features of Galaxy Zoo data (such as the beautiful galaxy images) or task (such as the fun of classifying galaxies); it was related to making a contribution to science in general.

More research is needed to understand how the motivations of Galaxy Zoo volunteers compare to the motivations of volunteers with other citizen science projects. However, the fact that most Galaxy Zoo users are motivated by a desire to contribute to science is encouraging for the future of citizen science.

## ACKNOWLEDGEMENTS

This publication has been made possible by the participation of more than 200,000 volunteers in the Galaxy Zoo project. Galaxy Zoo is supported by the Leverhulme Trust.

Support for the work of M.J.R., G.B., and P.L.G. was provided by NASA through ROSES EPOESS grant #NNX09AD34G.

Support for the work of K.S. was provided by NASA through Einstein Postdoctoral Fellowship grant number PF9-00069 Issued by the Chandra X-ray Observatory Center, which is operated by the Smithsonian Astrophysical Observatory for and on behalf of NASA under contract NAS8-03060.

# APPENDIX A

## Survey Instrument

Thank you very much for your interest in Galaxy Zoo. We very much appreciate the help you have given us in classifying galaxies – the work you have done helps us understand more about the universe that we all share.

We have been amazed by the skill and dedication of the thousands of volunteers who have given their time and skill to this project. Thank you. To help us to build further projects that will be like Galaxy Zoo, We are conducting an online survey to explore the motivations people have in working on Galaxy Zoo. The survey should take about 10 minutes to complete. Once you complete the survey, you will have beta-testing access to Galaxy Zoo 2.

## The Galaxy Zoo website

1. Which is more important to you – the enjoyment of classifying galaxies, or the knowledge that your classifications are being used for research?

1 2 3 4 5 6 7
enjoyment of classifying — classifications being used for research

2. On average, how difficult did you find it to classify galaxies when you first began?

1 2 3 4 5 6 7
very easy — very difficult

3. On average, how difficult do you find it to classify galaxies now?

1 2 3 4 5 6 7
very easy — very difficult

4. From where do you most often use Galaxy Zoo?
   a. From home
   b. From work
   c. From a computer in a public place (such as a library)
   d. Other:

## What are your reasons for participating in Galaxy Zoo?

5. Here are some reasons that other people have identified as reasons why they participate in Galaxy Zoo. How motivating are these reasons to you? Please choose a response to indicate how much of a motivating factor that reason is for you, on a scale from 1 (not at all motivating) to 7 (very motivating).

I am excited to contribute to original scientific research.
1 2 3 4 5 6 7
not motivating to me — very motivating to me

I find the site and forums helpful in learning about astronomy.
1 2 3 4 5 6 7
not motivating to me very motivating to me

I can look at galaxies that few people have ever seen before.
1 2 3 4 5 6 7
not motivating to me very motivating to me

I can meet other people with similar interests.
1 2 3 4 5 6 7
not motivating to me very motivating to me

I find Galaxy Zoo to be a useful resource for teaching other people.
1 2 3 4 5 6 7
not motivating to me very motivating to me

I enjoy looking at the beautiful galaxy images.
1 2 3 4 5 6 7
not motivating to me very motivating to me

I had a lot of fun categorizing the galaxies.
1 2 3 4 5 6 7
not motivating to me very motivating to me

I am amazed by the vast scale of the universe.
1 2 3 4 5 6 7
not motivating to me very motivating to me

I am happy to help.
1 2 3 4 5 6 7
not motivating to me very motivating to me

I am interested in the Galaxy Zoo project.
1 2 3 4 5 6 7
not motivating to me very motivating to me

I am interested in astronomy.

| 1 | 2 | 3 | 4 | 5 | 6 | 7 |
| --- | --- | --- | --- | --- | --- | --- |
| not motivating to me | | | | | | very motivating to me |

I am interested in science.

| 1 | 2 | 3 | 4 | 5 | 6 | 7 |
| --- | --- | --- | --- | --- | --- | --- |
| not motivating to me | | | | | | very motivating to me |

6. Can you think of any other reasons that someone might be interested in Galaxy Zoo? If so, list them here:

7. Which of the reasons above is MOST important to you for participating in Galaxy Zoo? Select the radio button next to your most important reason:

a. I am excited to contribute to original scientific research.
b. I find the site and forums helpful in learning about astronomy.
c. I can look at galaxies that few people have ever seen before.
d. I can meet other people with similar interests.
e. I find Galaxy Zoo to be a useful resource for teaching other people.
f. I enjoy looking at the beautiful galaxy images.
g. I had a lot of fun categorizing the galaxies.
h. I am amazed by the vast scale of the universe.
i. I am happy to help.
j. I am interested in the Galaxy Zoo project.
k. I am interested in astronomy.
l. I am interested in science.
m: Other:

## Demographics

8. What is your age?

9. What is your gender?

a. Male
b. Female
c. Prefer not to answer

10. In what country do you reside?

11. What is the highest level of education you have completed?

a. [These are the choices for participants in the U.S.:]

i. Less than high school
ii. Currently enrolled in high school
iii. High school diploma or GED
iv. Currently enrolled in college
v. Some college
vi. 2-year college (Associate or vocational) degree
vii. Bachelor’s degree (BA or BS)
viii. Currently enrolled in graduate program
ix. Some postgraduate study
x. Master’s degree (MA or MS)
xi. Doctoral degree (Ph.D., D.Phil…)
xii. Professional degree (M.D., J.D…)
xiii. Prefer not to answer
xiv. Other:

b. [These are the choices for participants in the U.K.:]

xv. Left school at 16 (GCSE / O levels)
xvi. Left school at 18 (e.g. A levels, highers, etc.)
xvii. Currently enrolled in secondary school
xviii. Currently enrolled in degree program
xix. First degree
xx. Currently enrolled in postgraduate study
xxi. Master's degree
xxii. Doctoral degree

d. [Add generic European system]

12. [U.S.:] What is your zip code? [U.K.:] What is the first part of your postcode?

13. Please check the box next to any science-related activities that you enjoy BEFORE and/or AFTER joining Galaxy Zoo?

____ Reading popular science magazines (such as *New Scientist* or *Scientific American)*
____ Reading popular science books
____ Amateur astronomy or stargazing
____ Other science hobbies (birdwatching, etc.)

____ Model building (model airplanes, rockets, etc.)
____ Reading or writing science fiction
____ Watching science television shows or specials
____ Other:

14. Please check the box next to any online collaboration projects you have participated in BEFORE and/or AFTER joining Galaxy Zoo:

____ SETI@Home
____ Stardust@Home
____ Wikipedia (as a reader)
____ Wikipedia (as a contributor)
____ Clickworkers
____ Amazon Mechanical Turk
____ The ESP game
____ Astronomy Picture of the Day
____ YouTube, Flickr or other media sites (As a viewer)
____ YouTube, Flickr or other media sites (As a contributor)
____ Cornell Birdwatching Public Science Project
Other:

## Attitudes toward science (optional)

15. Defining science is difficult because science is complex and does many things. But MAINLY science is (SELECT ONLY ONE):

a. a study of fields such as biology, chemistry, and physics
b. a body of knowledge, such as principles, laws and theories, which explain the world around us (matter, energy, and life).
c. exploring the unknown and discovering new things about our world and universe and how they work.
d. carrying out experiments to solve problems of interest in the world around us.
e. inventing or designing things (for example, artificial hearts, computers, space vehicles).
f. finding and using knowledge to make this world a better place to live in (for example, curing diseases, solving pollution, and improving agriculture).

g. an organization of people (called scientists) who have ideas and techniques for discovering new knowledge.
h. No one can define science.
i. I don't understand.
j. I don't know enough about this subject to make a choice.
k. None of these choices fits my basic viewpoint.